\documentclass[runningheads]{llncs}
\usepackage[T1]{fontenc}
\usepackage{graphicx}

\usepackage{hyperref}
\usepackage{bbm}
\usepackage{physics}
\usepackage[normalem]{ulem}
\usepackage{multirow}
\usepackage[table,xcdraw]{xcolor}
\usepackage{booktabs}
\usepackage{adjustbox}
\usepackage{calc}
\usepackage{amssymb}

\usepackage{color}

\begin{document}
\title{Enhancing Quantitative Image Synthesis\\
through Pretraining and Resolution Scaling\\
for Bone Mineral Density Estimation\\
from a Plain X-ray Image}
\titlerunning{Enhancing Quantitative Image Synthesis}
%
\author{
Yi Gu\inst{1} \and
Yoshito Otake\inst{1} \and
Keisuke Uemura\inst{2} \and
Masaki Takao\inst{3} \and
Mazen Soufi\inst{1} \and\\
Seiji Okada\inst{4} \and
Nobuhiko Sugano\inst{2} \and
Hugues Talbot\inst{5} \and
Yoshinobu Sato\inst{1}
}

%
\authorrunning{Y. Gu et al.}
%
\institute{
Division of Information Science, Graduate School of Science and Technology, \\
Nara Institute of Science and Technology, Japan\\
\email{gu.yi.gu4@naist.ac.jp, \{otake,yoshi\}@is.naist.jp} \and
Department of Orthopeadic Medical Engineering, \\
Osaka University Graduate School of Medicine, Japan \and
Department of Bone and Joint Surgery, \\
Ehime University Graduate School of Medicine, Japan \and
Department of Orthopaedics,
Osaka University Graduate School of Medicine, Japan \and
CentraleSupélec, Université Paris-Saclay, France
}
\maketitle              
\begin{abstract}
While most vision tasks are essentially visual in nature (for recognition), some important tasks, especially in the medical field, also require quantitative analysis (for quantification) using quantitative images.
Unlike in visual analysis, pixel values in quantitative images correspond to physical metrics measured by specific devices (e.g., a depth image). However, recent work has shown that it is sometimes possible to synthesize accurate quantitative values from visual ones (e.g., depth from visual cues or defocus). This research aims to improve quantitative image synthesis (QIS) by exploring pretraining and image resolution scaling.
We propose a benchmark for evaluating pretraining performance using the task of QIS-based bone mineral density (BMD) estimation from plain X-ray images, where the synthesized quantitative image is used to derive BMD.
Our results show that appropriate pretraining can improve QIS performance, significantly raising the correlation of BMD estimation from 0.820 to 0.898, while others do not help or even hinder it. 
Scaling-up the resolution can further boost the correlation up to 0.923, a significant enhancement over conventional methods. Future work will include exploring more pretraining strategies and validating them on other image synthesis tasks.
\end{abstract}
\begin{figure}
\includegraphics[width=\textwidth]{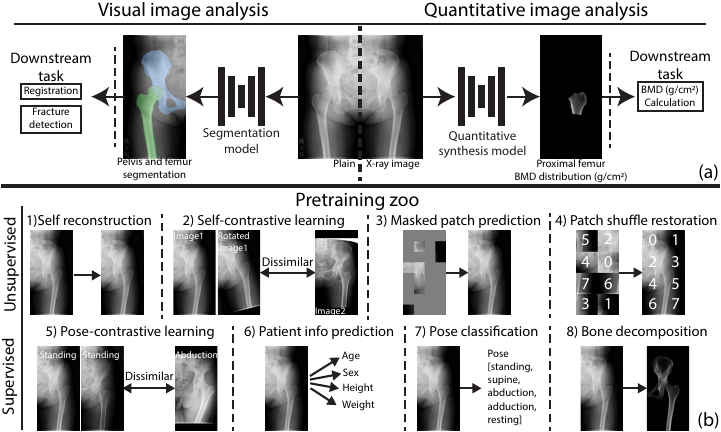}
\caption{(a) A comparison between visual and quantitative image analysis. (b) Our pretraining zoo consists of several unsupervised and supervised pretraining.} \label{fig:overview}
\end{figure}

\section{Introduction}
Image synthesis is a popular research direction that provides critical support for many medical applications \cite{dayarathna_deep_2024,osuala_data_2023,wang_spatial-intensity_2023,azampour_multitask_2024,vorontsov_towards_2022,gu_bmd-gan_2022,gu_mskdex_2023,gu_bone_2023}. 
There are two broad categories in image analysis \cite{buvat_quantitative_2012}: 1) \textit{quantitative} image analysis, where objective and quantitative information is derived from pixel values representing measurements, and 2) \textit{visual} image analysis, which derives information from pixel relations. 
Fig. \ref{fig:overview} (a) shows two visual and quantitative image analysis examples.
Although most vision tasks rely on visual image analysis \cite{esteva_deep_2021,cheng_fashion_2021,hoye_deep_2021,lu_survey_2020,noothout_deep_2020}, some important tasks, especially in the medical field, require quantitative image analysis \cite{gu_bone_2023,gu_bmd-gan_2022,gu_mskdex_2023}. 
Image synthesis can also be divided into quantitative image synthesis (QIS) and visual image synthesis according to the downstream tasks.
We focus on improving QIS performance for its high-potential quantification application in the medical field. 
Pretraining is a widely used technique to improve deep-learning models \cite{carr_self-supervised_2021,wang_contrastive_2022,khosla_supervised_2020,chen_simple_2020,he_momentum_2020,he_masked_2022}.
However, it has barely been explored for QIS. In this article, We compared the effects of various pretrainings on a QIS task, where a quantitative image of the proximal femur (PF) is synthesized from a plain X-ray and then used to derive bone mineral density (BMD) \cite{gu_bmd-gan_2022,gu_bone_2023}, an important metric to determine osteoporosis.
Fig. \ref{fig:overview} (b) summarizes our pretraining zoo.
Our results show that appropriate pretraining can improve BMD estimation performance, while others may have degraded it.
By combining the best pretraining strategy with resolution scaling, we significantly enhanced the conventional methods by 13\%.
Our code is available at \href{https://github.com/Kayaba-Akihiko/next-bmdgan-tutorial}{https://github.com/Kayaba-Akihiko/next-bmdgan-tutorial}.

\noindent \textbf{Contributions:}  
1) a pretraining zoo of 10 techniques over a large X-ray image database.
2) A large-scale benchmark on a 2656-X-ray dataset for validating the pretraining effect on the QIS-based BMD estimation task.
3) Finding that our best pretraining combined with the highest resolution significantly enhances QIS performance and BMD estimation accuracy.

\section{Related Work}
\textbf{Image Synthesis} and translation aim at image generation, usually serving downstream tasks or pipelines. 
The final performance is heavily dependent on image synthesis performance.
\cite{azampour_multitask_2024} used image translation of magnetic resonance to ultrasound to improve registration accuracy.
\cite{vorontsov_towards_2022,gilbert_generating_2021} used synthesized images to train segmentation models.
These previous works addressed visual image synthesis in which the error of the absolute pixel intensity does not affect the downstream tasks.
On the other hand, \cite{gu_bmd-gan_2022,gu_mskdex_2023} utilized QIS for BMD estimation and lean muscle mass prediction, in which the physical metrics were derived/calculated from synthesized quantitative images.

\noindent \textbf{Pretraining} learns representations from source datasets and tasks to improve the performance on the target datasets and tasks \cite{he_masked_2022,he_momentum_2020,chen_simple_2020,khosla_supervised_2020,wang_contrastive_2022,carr_self-supervised_2021}.
Contrastive learning is often used to learn representations \cite{hadsell_dimensionality_2006}.
Self-supervised pretraining \cite{carr_self-supervised_2021,chen_simple_2020,he_momentum_2020,he_masked_2022} requiring no annotation for the source datasets is popular in this field due to its label-free nature and high generalizability.
On the other hand, supervised pretraining (with specialized datasets and tasks) \cite{wang_contrastive_2022,khosla_supervised_2020} exhibit better performance gain than self-supervised ones; however, the availability of labels and the low generalizability limit applicability.
Our research compares several pretrainings to improve the QIS-based BMD estimation task \cite{gu_bmd-gan_2022,gu_bone_2023}.

\noindent \textbf{BMD estimation from plain X-ray} has drawn wide attention recently due to its high potential for fast and large-scale osteoporosis screening, an important public health problem worldwide \cite{hsieh_automated_2021,gu_bmd-gan_2022,gu_bone_2023,wang_lumbar_2023,ho_application_2021,nguyen_enhancement_2023}.
A straightforward way to estimate BMD is to directly regress it from large-scale training datasets \cite{hsieh_automated_2021,wang_lumbar_2023,ho_application_2021,nguyen_enhancement_2023}.
However the regression method's low explainability and large dataset requirement limit applications.
To address that, \cite{gu_bmd-gan_2022,gu_bone_2023} proposed a QIS-based method that estimates BMD distribution map (a quantitative image derived from CT data) from an X-ray to calculate BMD, achieving high correlation and explainability from smaller datasets.
We show that BMD estimation performance can be further improved with appropriate pretraining and resolution scaling.

\section{Dataset}
Our database includes 600 patients who underwent 4 or 5  X-ray scans with varying poses (standing, supine, resting, abduction, and adduction) and a single CT scan with a phantom \cite{uemura_automated_2021} with known densities of hydroxyapatite $\mathrm{Ca_{10}(PO_{4})_{6}(OH)_{2}}$ [to calibrate the quantitative CT (QCT)] at the same hospital. 
Each patient's profile contained age, sex, height, and weight.
We followed \cite{gu_bone_2023} for dataset construction.
The BMD distribution maps of the bone and PF (the bone DRR and PF-DRR in the \cite{gu_bone_2023}, respectively) were obtained by projecting masked QCT \cite{hiasa_automated_2020} after 2D-3D registration \cite{otake_intraoperative_2012} to the X-ray.
The X-ray (and the corresponding BMD distribution maps) was split into left and right halves.
For the BMD estimation task, only one-side X-ray was measured.
For pretraining, both sides are used.
We obtained 2656 pairs of an X-ray and a BMD distribution map of the PF region for the target task and 5312 pairs of an X-ray and a BMD distribution map of the bone region for pretraining.

\section{Method}
We constructed a generator $G$ consisting of an encoder and a decoder.
The decoder takes the feature pyramid from the encoder to produce the target image.
When pretraining is applied, the encoder is initialized with the weights from the pretrained model instead of random weights.
We present the details of our pretraining tasks and the BMD estimation task in Sec. \ref{sec:method:pretrain} and Sec. \ref{sec:method:bmd}, respectively.
The pretraining and the BMD estimation tasks share the same database.
In addition to exploring pretrainings, we investigated the performance of varying X-ray resolutions, which we report in Sec. \ref{sec:experiment}.

\subsection{Pretraining zoo}
\label{sec:method:pretrain}
We grouped the pretraining $\mathbf{P}_i$ into self-supervised ($\mathbf{P_{sr}}$, $\mathbf{P_{mp}}$, $\mathbf{P_{ps}}$, and $\mathbf{P_{sc}}$) and supervised ones ($\mathbf{P_{pac}}$, $\mathbf{P_{poc}}$, $\mathbf{P_{pc}}$, $\mathbf{P_{pi}}$, and $\mathbf{P_{bd}}$) in the following sections.
We denote $\mathbf{P_0}$ as the baseline that used no pretraining.
Inspired by the Hierarchical PreTraining (HPT) \cite{reed_self-supervised_2022}, we also adopted cascaded pretraining denoted by $\mathbf{P}_{i,j}$.
We used the MoCo trick \cite{he_momentum_2020} for contrastive learning.
Though most of our pretraining tasks were based on visual image analysis, the $\mathbf{P_{bd}}$ from \cite{gu_bone_2023} is a specialized task that uses quantitative images.

\noindent \textbf{$\mathbf{P_{sr}}$ Self reconstruction} is a classic task that uses an autoencoder to learn representations by reconstructing the input image.
We adopted an asymmetric autoencoder to learn representation from X-ray images, where the decoder is lightweight.
We used the mean squared error (MSE) between the input and reconstructed X-rays as the loss function.

\noindent \textbf{$\mathbf{P_{mp}}$ Masked patch prediction} learns representation by predicting the masked patches from an image in a self-supervised fashion.
We follow most settings from masked autoencoders \cite{he_masked_2022} that divide an image (an X-ray image in our case) into patches and used a masking ratio of 75\%.
Unlike the masked autoencoder, we fed both the masked and unmasked patches to the encoder (which was, though, inefficient) to adopt convolutional neural network-based encoders.

\noindent \textbf{$\mathbf{P_{ps}}$ Patch shuffle restoration} tries to reorder the shuffled image patches to learn representation.
We borrowed the shuffle idea from \cite{carr_self-supervised_2021} and applied it to X-ray image patches.
For simplicity, we used a cross-entropy loss instead of the differentiable ranking loss \cite{carr_self-supervised_2021}.
The sorting problem was then converted into a classification problem, where the model was required to predict the correct index for each patch.
Thus, our loss function $\mathcal{L}_s$ is defined as $\mathcal{L}_s=\sum_i^K\sum_j^K-y_{i,j}\mathrm{log}p_{i,j}$, where the $p_{i,j}$ is the predicted probability of the $i$-th patch in the input image to be $j$-th patch given $K$ patches; $y_{i,j}$ is the corresponding label.

\noindent \textbf{$\mathbf{P_{sc}}$ Self-supervised contrastive learning} on X-ray images uses an unsupervised contrastive loss to cluster the similar representations (encoded by the encoder).
We used the InfoNCE loss \cite{oord_representation_2019} $-\mathrm{log}\frac{\mathrm{exp}(q\cdot k_{+}/\tau)}{\sum_i^K\mathrm{exp}(q\cdot k_i/\tau)}$ for similarity measurement, where the $q$, $k_+$, and $k_i$ are the encoded query, positive sample, and $i$-th of $K$ sample, respectively.
where $\tau$ is temperature hyperparameter.

\noindent \textbf{$\mathbf{P_{pac}}$ Patient-supervised contrastive learning} leverages the power of supervised contrastive learning loss~\cite{khosla_supervised_2020} that normalizes the self-supervised contrastive learning loss by the fact that the multiple positive samples (belong to the same class) are presented. 
The loss is defined as $\frac{-1}{N}\sum_i^N\mathrm{log}\frac{\mathrm{exp}(q\cdot k_i/\tau)}{\sum_j^K\mathrm{exp}(q\cdot k_j/\tau)}$, where $k_i$ represents $i$-th of $N$ positive samples to a query $q$; $k_j$ means $j$-th of $K$ samples.
This task considered the X-ray images of the same patient (in different poses) to be of the same class.

\noindent \textbf{$\mathbf{P_{poc}}$ Pose-supervised contrastive learning} is similar to patient-supervised contrastive learning $\mathbf{P_{pac}}$, where the X-ray images of the same pose (from different patients) were considered to be the same class.

\noindent \textbf{$\mathbf{P_{pc}}$ Pose classification} is a task to classify the patient pose given an X-ray. 
This is a $K$-classes, $K=5$ classification problem in our dataset.
A linear classifier was attached to the encoder to perform classification.
We used the cross-entropy loss $\mathcal{L}_{ce}=\sum_i^K-y_i\log p_i$, where the $p_i$ is the predicted probability of $i$-th class; $y_i$ is the corresponding ground-truth label of the $i$-th class.

\noindent \textbf{$\mathbf{P_{pi}}$ Patient info prediction} expects a model to predict the patient information of age $a$, sex $s$ (male or female), height $h$, and weight $w$ from a given X-ray.
We attached a linear classifier that performs classification and regression simultaneously, where a mix of cross-entropy loss $\mathcal{L}_{bce}=-s\log \hat{s}_p-(1-s)\log(1-\hat{s}_p)$ and squared error loss $\mathcal{L}_{se}=(a-\hat{a})^2+(h-\hat{h})^2+(w-\hat{w})^2$ was used.
The $\hat{a}$, $\hat{h}$, $\hat{w}$, and $\hat{s}_p$ represent the predicted age, height, weight, and sex probability. The final loss in this pretraining task is defined as $\mathcal{L}=\mathcal{L}_{bce}+\mathcal{L}_{se}$.

\noindent \textbf{$\mathbf{P_{bd}}$ Bone decomposition from X-ray} was mentioned in \cite{gu_bmd-gan_2022,gu_bone_2023} as the first stage/task of a hierarchical learning scheme, where the model was trained to generate a BMD distribution map of the pelvis and femur bone from an X-ray.
We now summarize hierarchical learning as the use of specialized supervised pretraining.
For simplicity and better alignment with other pretraining tasks, we only used the MSE loss.

\noindent \textbf{$\mathbf{P_{mp,bd}}$ Sequential pretraining of $\mathbf{P_{mp}}$ and $\mathbf{P_{bd}}$} uses the idea of HPT \cite{reed_self-supervised_2022} to initialize the pretraining with a base self-supervised pretraining.
We selected two promising pretraining, $\mathbf{P_{mp}}$ and $\mathbf{P_{bd}}$, to form the HPT framework, where the $\mathbf{P_{mp}}$ was the base training to improve $\mathbf{P_{bd}}$.

\subsection{BMD estimation task}
\label{sec:method:bmd}
The target task \cite{gu_bmd-gan_2022,gu_bone_2023} expects a generator $G$ to synthesize a BMD distribution map $G(x)$ of the PF region (defined clinically) from an X-ray $x$ to calculate the BMD.
Following \cite{gu_bmd-gan_2022,gu_bone_2023}, we trained a generative adversarial network (GAN) with a generator $G$ and a discriminator $D$.
The adversarial loss $\mathcal{L}_\mathrm{GAN}$ is defined as
\begin{equation}
    \mathcal{L}_\mathrm{GAN}=\mathbbm{E}_{y}[\log D(y)]+\mathbbm{E}_{x}[\log (1-D(G(x)))],
\end{equation}
where $x$ and $y$ are the X-ray image and BMD distribution map, respectively \footnote{We denote $\mathbbm{E}_{x}\triangleq\mathbbm{E}_{x\sim p_{data}(x)}$, $\mathbbm{E}_{y}\triangleq\mathbbm{E}_{y\sim p_{data}(y)}$, and $\mathbbm{E}_{x,y}\triangleq\mathbbm{E}_{x,y\sim p_{data}(x,y)}$}.
Though the L2 loss $\mathcal{L}_{\mathrm{L2}}=\mathbbm{E}_{x,y}[\norm{y-G(x)}_2]$ causes more blurring than the L1 loss, we found that it provides a more accurate quantitative estimation.
We removed the feature matching loss and multi-discriminators since we found them less effective in this task.
However, we did keep the gradient correlation (GC) loss defined as $\mathcal{L}_{GC}=\mathbbm{E}_{x,y}[-GC(y,G(x))]$, where $GC(\cdot)$ calculates the gradient correlation between two images.
The full objective is defined as $\min_G\left(\lambda_{L2}\mathcal{L}_{L2}+\lambda_{GC}\mathcal{L}_{GC}+\max_D\mathcal{L}_{GAN}\right)$,
where the $\lambda_{L2}$ and $\lambda_{GC}$ are the balance hyper-parameters.
During inference, the estimated BMD $b$ was calculated from the estimated BMD distribution map $G(x)$ by Eq. \ref{eq:bmd}, where $G(x)_i$ represents $i$-th pixels of $G(x)$; $th$ is a threshold to determine PF region \cite{gu_bone_2023}.
\begin{equation}
\label{eq:bmd}
    b=\frac{\sum_i\mathbbm{1}\{G(x)_i > th\}G(x)_i}{\sum_i\mathbbm{1}\{G(x)_i > th\}}
\end{equation}

\begin{figure}
\centering
\includegraphics[width=\textwidth]{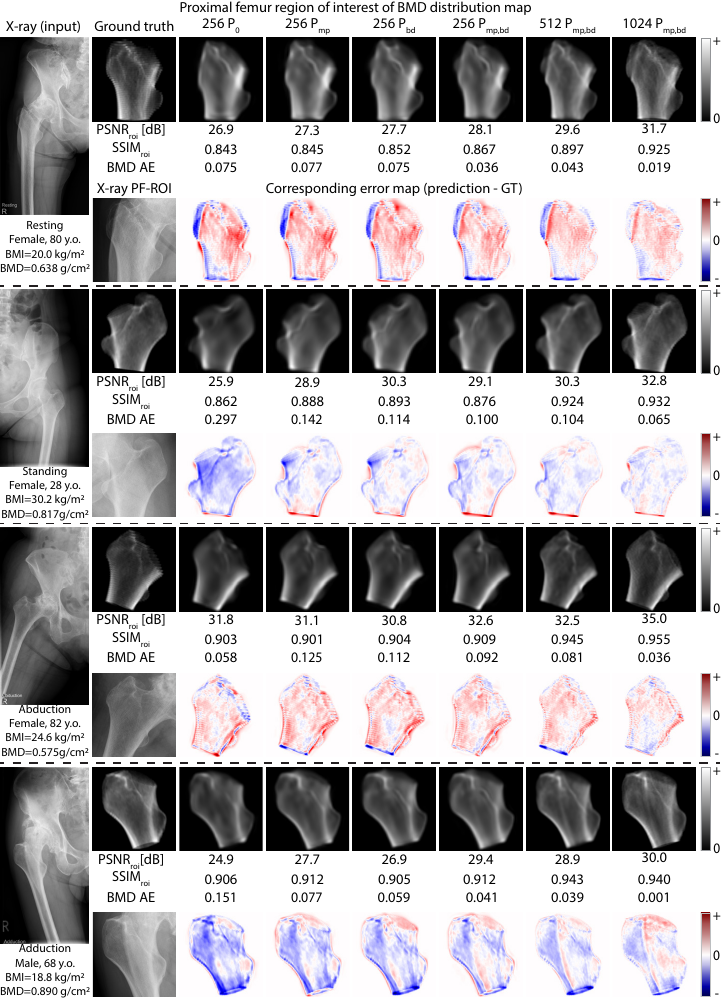}
\caption{Qualitative comparison of PF-BMD distribution map prediction with different pretrain strategies.} \label{fig:visual_pred}
\end{figure}

\section{Experiment and Result}
\label{sec:experiment}
We applied a patient-wise four-fold cross-validation on our X-ray dataset.
The pretraining and the target training share the same training dataset under the same cross-validation policy.
We used the QCT-aBMD ($\mathrm{g/cm^2}$) derived from QCT as the BMD ground truth (GT).
We explored the pretraining effect first on a downsampled resolution of 256 (which represents the image size of 128$\times$256, width$\times$height of a half X-ray) and then increased it to 1024.
To assess the QIS performance, we evaluated 1) BMD estimation accuracy and 2) image quality of the synthesized BMD distribution map.
For 1), we used the Pearson correlation coefficient (PCC), Intra-class correlation coefficient (ICC), and mean absolute error (mAE).
For 2), we used Peak-signal-to-noise-ratio (PSNR) and structural similarity index measure (SSIM) and reported their mean(std) results denoted by mPSNR and mSSIM, respectively.
The estimated BMD distribution map was restored to the original resolution of its X-ray before evaluation.
We also evaluate the PF region of interest (ROI) in the estimated PF BMD distribution map with PSNR and SSIM denoted as $\mathrm{PSNR_{roi}}$ and $\mathrm{SSIM_{roi}}$, respectively.
Implementation details are summarized in supplementary materials.

\noindent \textbf{Qualitative evaluation} for four representative X-rays (of different patients and poses) are shown in Fig. \ref{fig:visual_pred}, where the 256, 512, and 1024 denote the X-ray resolutions.
The error maps indicate the over- (red) and under- (blue) predicted regions.
Compared to training from scratch (256 $\mathbf{P_0}$), pretraining (256 $\mathbf{P_{mp,bd}}$) improved the visual quality of the predicted PF-BMD distribution map, allowing it to align better with the GT indicated by the error map.
The increased resolution (1024 $\mathbf{P_{mp,bd}}$) further improved the visual quality significantly, making the synthesized texture of PF much clearer to achieve higher scores of PSNR and SSIM, demonstrating a strong enhancement on the synthesis performance. 
\noindent \textbf{Quantitative evaluation} of our pretraining is summarized in Tab. \ref{tab:eval}.
The pretraining under 256 resolution improved image evaluation metrics steadily from 28.6 and 0.887 to 29.9 and 0.898 for the $\mathrm{PSNR_{roi}}$ and $\mathrm{SSIM_{roi}}$, respectively.
Furthermore, the cascaded pretraining ($\mathbf{P_{mp,bd}}$) improved the previously proposed decomposition pretraining \cite{gu_bone_2023} ($\mathbf{P_{bd}}$).
The highest resolution of the best pretraining strategy (1024 $\mathbf{P_{mp,bd}}$), achieved the best scores of 31.0 and 0.928 for $\mathrm{PSNR_{roi}}$ and $\mathrm{SSIM_{roi}}$, respectively, a large improvement over $\mathbf{P_0}$.
Interestingly, some pretrainings appeared harmful to the downstream task of BMD estimation.
For example, the self-contrastive learning ($\mathbf{P_{sc}}$) and pose classification ($\mathbf{P_{pc}}$) degraded the PCC from 0.820 to 0.811 and 0.804, respectively.
The embedded pose text in the X-ray (Fig. \ref{fig:visual_pred}) may have misled the model to learn unrelated features during pretraining, which degraded the performance for the target task.
The best pretraining with the highest resolution (1024 $\mathbf{P_{mp, bd}}$) achieved a high PCC of 0.923 and significantly improved the ICC from 0.812 to 0.914.
{
\DeclareFontSeriesDefault[rm]{bf}{sb}   
\begin{table}[]
\centering
\caption{Quantitative evaluation summary}
\label{tab:eval}
\begin{tabular}{cccccccc}
\toprule
\hline
                            & \multicolumn{1}{c|}{}                              & \multicolumn{3}{c|}{\textbf{BMD evaluation}}      & \multicolumn{3}{c}{\textbf{Image evaluation}$\uparrow$($\downarrow$)}   \\
\multirow{-2}{*}{\textbf{Res.}}      & \multicolumn{1}{c|}{\multirow{-2}{*}{\textbf{Pretraining}}} & PCC$\uparrow$   & ICC$\uparrow$   & \multicolumn{1}{c|}{mAE$\downarrow$($\downarrow$)} & mPSNR      & $\mathrm{mPSNR_{roi}}$   & $\mathrm{mSSIM_{roi}}$     \\ \hline
256                         & $\mathbf{P_0}$                                                 & 0.820 & 0.812 & 0.069(0.055)             & 39.9(2.52) & 28.6(\textbf{2.27}) & 0.887(0.031) \\
{\color[HTML]{C0C0C0} 256}  & $\mathbf{P_{sr}}$                                                  & 0.845 & 0.842 & 0.062(0.050)             & 40.2(2.52) & 28.9(2.28) & 0.890(0.030) \\
{\color[HTML]{C0C0C0} 256}  & $\mathbf{P_{mp}}$                                                  & 0.869 & 0.864 & 0.056(0.045)             & 40.7(2.54) & 29.4(2.31) & 0.894(0.029) \\
{\color[HTML]{C0C0C0} 256}  & $\mathbf{P_{ps}}$                                                  & 0.856 & 0.825 & 0.067(0.051)             & 40.6(\textbf{2.51}) & 29.3(2.27) & 0.892(0.029) \\
{\color[HTML]{C0C0C0} 256}  & $\mathbf{P_{sc}}$                                                  & 0.811 & 0.797 & 0.066(0.054)             & 40.3(2.62) & 29.0(2.39) & 0.890(0.030) \\
{\color[HTML]{C0C0C0} 256}  & $\mathbf{P_{pac}}$                                                  & 0.827 & 0.814 & 0.064(0.052)             & 40.4(2.63) & 29.0(2.39) & 0.892(0.030) \\
{\color[HTML]{C0C0C0} 256}  & $\mathbf{P_{poc}}$                                                  & 0.823 & 0.779 & 0.074(0.055)             & 40.4(2.53) & 29.1(2.30) & 0.891(0.030) \\
{\color[HTML]{C0C0C0} 256}  & $\mathbf{P_{pc}}$                                                  & 0.804 & 0.797 & 0.071(0.056)             & 40.2(2.57) & 28.9(2.33) & 0.892(0.030) \\
{\color[HTML]{C0C0C0} 256}  & $\mathbf{P_{pi}}$                                                  & 0.822 & 0.811 & 0.067(0.053)             & 40.4(2.56) & 29.0(2.32) & 0.892(0.030) \\
{\color[HTML]{C0C0C0} 256}  & $\mathbf{P_{bd}}$                                                  & 0.886 & 0.876 & 0.053(\textbf{0.042})             & 40.9(2.53) & 29.6(2.29) & 0.897(0.029) \\
{\color[HTML]{C0C0C0} 256}  & $\mathbf{P_{mp,bd}}$                                                 & {\textbf{0.898}} & \textbf{0.886} & \textbf{0.049}(\textbf{0.042})             & \textbf{41.3}(2.53) & \textbf{29.9}(2.32) & \textbf{0.898}(\textbf{0.028}) \\ \hline
512                         & $\mathbf{P_{0}}$                                                  & 0.827 & 0.758 & 0.079(0.063)             & 41.2(2.86) & 29.9(2.60) & 0.920(0.031) \\
{\color[HTML]{C0C0C0} 512}  & $\mathbf{P_{bd}}$                                                  & 0.905 & 0.892 & 0.049(0.040)             & \textbf{42.3}(2.93) & \textbf{30.9}(2.66) & 0.926(\textbf{0.029})  \\
{\color[HTML]{C0C0C0} 512}  & $\mathbf{P_{mp,bd}}$                                                 & \textbf{0.917} & \textbf{0.909} & \textbf{0.044}(\textbf{0.038})             & \textbf{42.3}(\textbf{2.80}) & \textbf{30.9}(\textbf{2.54}) & \textbf{0.925}(\textbf{0.029}) \\ \hline
1024                        & $\mathbf{P_{0}}$                                                  & 0.878 & 0.869 & 0.054(0.045)             & 41.1(\textbf{2.73}) & 29.7(\textbf{2.47}) & 0.920(0.031) \\
{\color[HTML]{C0C0C0} 1024} & $\mathbf{P_{mp,bd}}$                                                 & \textbf{0.923} & \textbf{0.914} & \textbf{0.043}(\textbf{0.036})             & \textbf{42.3}(2.75) & \textbf{31.0}(\textbf{2.47}) & \textbf{0.928}(\textbf{0.029}) \\ \hline
\end{tabular}
\end{table}
}

\section{Conclusion}
We demonstrated improvement to the QIS performance by applying pretraining and scaling up the resolution, taking QIS-based BMD estimation as an instance target.
We built a pretraining zoo with 10 strategies and applied them to the target BMD estimation task. Through evaluation on a 2656-X-ray dataset, all the pretraining improved image-based metrics such as PSNR and SSIM. However, some pretraining hurt BMD estimation performance.
Our results suggest that one must carefully choose the pretraining (in addition to choosing source datasets \cite{reed_self-supervised_2022}) when targeting QIS-based tasks. Future work includes exploring other pretraining and validation on other QIS-based tasks with more medical datasets.

\begin{credits}
\subsubsection{\ackname} The research in this paper was funded by\\
MEXT/JSPS KAKENHI (19H01176, 20H04550, 21K16655).
\subsubsection{\discintname}
The authors have no competing interests to declare relevant to this article's content.
\end{credits}

%
%
%
\bibliographystyle{splncs04}
\bibliography{Paper-030}

\begin{thebibliography}{10}
\providecommand{\url}[1]{\texttt{#1}}
\providecommand{\urlprefix}{URL }
\providecommand{\doi}[1]{https://doi.org/#1}

\bibitem{azampour_multitask_2024}
Azampour, M.F., et~al.: Multitask {Weakly} {Supervised} {Generative} {Network} for {MR}-{US} {Registration}. IEEE Transactions on Medical Imaging pp.~1--1 (2024)

\bibitem{buvat_quantitative_2012}
Buvat, I.: Quantitative {Image} {Analysis} in {Tomography}. In: Grupen, C., Buvat, I. (eds.) Handbook of {Particle} {Detection} and {Imaging}, pp. 1043--1063. Springer, Berlin, Heidelberg (2012). \doi{10.1007/978-3-642-13271-1_41}

\bibitem{carr_self-supervised_2021}
Carr, A.N., et~al.: Self-{Supervised} {Learning} of {Audio} {Representations} {From} {Permutations} {With} {Differentiable} {Ranking}. IEEE Signal Processing Letters  \textbf{28},  708--712 (2021). \doi{10.1109/LSP.2021.3067635}

\bibitem{chen_simple_2020}
Chen, T., et~al.: A {Simple} {Framework} for {Contrastive} {Learning} of {Visual} {Representations}. In: Proceedings of the 37th {International} {Conference} on {Machine} {Learning}. pp. 1597--1607. PMLR (Nov 2020)

\bibitem{cheng_fashion_2021}
Cheng, W.H., Song, S., Chen, C.Y., Hidayati, S.C., Liu, J.: Fashion {Meets} {Computer} {Vision}: {A} {Survey}. ACM Comput. Surv.  \textbf{54}(4),  72:1--72:41 (Jul 2021)

\bibitem{dayarathna_deep_2024}
Dayarathna, S., et~al.: Deep learning based synthesis of {MRI}, {CT} and {PET}: {Review} and analysis. Medical Image Analysis  \textbf{92},  103046 (Feb 2024)

\bibitem{esteva_deep_2021}
Esteva, A., et~al.: Deep learning-enabled medical computer vision. npj Digit. Med.  \textbf{4}(1), ~1--9 (Jan 2021). \doi{10.1038/s41746-020-00376-2}

\bibitem{gilbert_generating_2021}
Gilbert, A., et~al.: Generating {Synthetic} {Labeled} {Data} {From} {Existing} {Anatomical} {Models}: {An} {Example} {With} {Echocardiography} {Segmentation}. IEEE Transactions on Medical Imaging  \textbf{40}(10),  2783--2794 (Oct 2021)

\bibitem{gu_bmd-gan_2022}
Gu, Y., et~al.: {BMD}-{GAN}: {Bone} {Mineral} {Density} {Estimation} {Using} {X}-{Ray} {Image} {Decomposition} into {Projections} of {Bone}-{Segmented} {Quantitative} {Computed} {Tomography} {Using} {Hierarchical} {Learning}. In: Wang, L., Dou, Q., Fletcher, P.T., Speidel, S., Li, S. (eds.) Medical {Image} {Computing} and {Computer} {Assisted} {Intervention} – {MICCAI} 2022. pp. 644--654. Springer Nature Switzerland, Cham (2022)

\bibitem{gu_bone_2023}
Gu, Y., et~al.: Bone mineral density estimation from a plain {X}-ray image by learning decomposition into projections of bone-segmented computed tomography. Medical Image Analysis  \textbf{90},  102970 (Dec 2023)

\bibitem{gu_mskdex_2023}
Gu, Y., et~al.: {MSKdeX}: {Musculoskeletal} ({MSK}) {Decomposition} from an {X}-{Ray} {Image} for {Fine}-{Grained} {Estimation} of {Lean} {Muscle} {Mass} and {Muscle} {Volume}. In: Greenspan, H., Madabhushi, A., Mousavi, P., Salcudean, S., Duncan, J., Syeda-Mahmood, T., Taylor, R. (eds.) Medical {Image} {Computing} and {Computer} {Assisted} {Intervention} – {MICCAI} 2023. pp. 497--507. Springer Nature Switzerland, Cham (2023). \doi{10.1007/978-3-031-43990-2_47}

\bibitem{hadsell_dimensionality_2006}
Hadsell, R., et~al.: Dimensionality {Reduction} by {Learning} an {Invariant} {Mapping}. In: 2006 {IEEE} {Computer} {Society} {Conference} on {Computer} {Vision} and {Pattern} {Recognition} ({CVPR}'06). vol.~2, pp. 1735--1742 (Jun 2006)

\bibitem{he_momentum_2020}
He, K., Fan, o.: Momentum {Contrast} for {Unsupervised} {Visual} {Representation} {Learning}. In: 2020 {IEEE}/{CVF} {Conference} on {Computer} {Vision} and {Pattern} {Recognition} ({CVPR}). pp. 9726--9735. IEEE, Seattle, WA, USA (Jun 2020)

\bibitem{he_masked_2022}
He, K., et~al.: Masked {Autoencoders} {Are} {Scalable} {Vision} {Learners}. In: 2022 {IEEE}/{CVF} {Conference} on {Computer} {Vision} and {Pattern} {Recognition} ({CVPR}). pp. 15979--15988. IEEE, New Orleans, LA, USA (Jun 2022)

\bibitem{hiasa_automated_2020}
Hiasa, Y., et~al.: Automated {Muscle} {Segmentation} from {Clinical} {CT} {Using} {Bayesian} {U}-{Net} for {Personalized} {Musculoskeletal} {Modeling}. IEEE Transactions on Medical Imaging  \textbf{39}(4),  1030--1040 (Apr 2020). \doi{10.1109/TMI.2019.2940555}

\bibitem{ho_application_2021}
Ho, C.S., et~al.: Application of deep learning neural network in predicting bone mineral density from plain {X}-ray radiography. Arch Osteoporos  \textbf{16}(1), ~153 (Oct 2021). \doi{10.1007/s11657-021-00985-8}

\bibitem{hsieh_automated_2021}
Hsieh, C.I., et~al.: Automated bone mineral density prediction and fracture risk assessment using plain radiographs via deep learning. Nat Commun  \textbf{12}(1), ~5472 (Sep 2021). \doi{10.1038/s41467-021-25779-x}

\bibitem{hoye_deep_2021}
Høye, T.T., et~al.: Deep learning and computer vision will transform entomology. Proceedings of the National Academy of Sciences  \textbf{118}(2),  e2002545117 (Jan 2021)

\bibitem{khosla_supervised_2020}
Khosla, P., et~al.: Supervised {Contrastive} {Learning}. In: Advances in {Neural} {Information} {Processing} {Systems}. vol.~33, pp. 18661--18673. Curran Associates, Inc. (2020)

\bibitem{lu_survey_2020}
Lu, Y., Young, S.: A survey of public datasets for computer vision tasks in precision agriculture. Computers and Electronics in Agriculture  \textbf{178},  105760 (Nov 2020)

\bibitem{nguyen_enhancement_2023}
Nguyen, T.P., et~al.: Enhancement of {Hip} {X}-ray with {Convolutional} {Autoencoder} for {Increasing} {Prediction} {Accuracy} of {Bone} {Mineral} {Density}. Bioengineering  \textbf{10}(10), ~1169 (Oct 2023)

\bibitem{noothout_deep_2020}
Noothout, J.M.H., et~al.: Deep {Learning}-{Based} {Regression} and {Classification} for {Automatic} {Landmark} {Localization} in {Medical} {Images}. IEEE Transactions on Medical Imaging  \textbf{39}(12),  4011--4022 (Dec 2020)

\bibitem{oord_representation_2019}
Oord, A.v.d., et~al.: Representation {Learning} with {Contrastive} {Predictive} {Coding} (Jan 2019). \doi{10.48550/arXiv.1807.03748}

\bibitem{osuala_data_2023}
Osuala, R., et~al.: Data synthesis and adversarial networks: {A} review and meta-analysis in cancer imaging. Medical Image Analysis  \textbf{84},  102704 (Feb 2023)

\bibitem{otake_intraoperative_2012}
Otake, Y., et~al.: Intraoperative image-based multiview {2D}/{3D} registration for image-guided orthopaedic surgery: incorporation of fiducial-based {C}-arm tracking and {GPU}-acceleration. IEEE Trans Med Imaging  \textbf{31}(4),  948--962 (Apr 2012)

\bibitem{reed_self-supervised_2022}
Reed, C.J., et~al.: Self-{Supervised} {Pretraining} {Improves} {Self}-{Supervised} {Pretraining}. In: 2022 {IEEE}/{CVF} {Winter} {Conference} on {Applications} of {Computer} {Vision} ({WACV}). pp. 1050--1060. IEEE, Waikoloa, HI, USA (Jan 2022)

\bibitem{uemura_automated_2021}
Uemura, K., et~al.: Automated segmentation of an intensity calibration phantom in clinical {CT} images using a convolutional neural network. Int J Comput Assist Radiol Surg  \textbf{16}(11),  1855--1864 (Nov 2021)

\bibitem{vorontsov_towards_2022}
Vorontsov, E., et~al.: Towards annotation-efficient segmentation via image-to-image translation. Medical Image Analysis  \textbf{82},  102624 (Nov 2022)

\bibitem{wang_spatial-intensity_2023}
Wang, C.J., et~al.: Spatial-{Intensity} {Transforms} for {Medical} {Image}-to-{Image} {Translation}. IEEE Transactions on Medical Imaging  \textbf{42}(11),  3362--3373 (Nov 2023)

\bibitem{wang_lumbar_2023}
Wang, F., et~al.: Lumbar {Bone} {Mineral} {Density} {Estimation} {From} {Chest} {X}-{Ray} {Images}: {Anatomy}-{Aware} {Attentive} {Multi}-{ROI} {Modeling}. IEEE Transactions on Medical Imaging  \textbf{42}(1),  257--267 (Jan 2023)

\bibitem{wang_contrastive_2022}
Wang, Y., et~al.: Contrastive {Regression} for {Domain} {Adaptation} on {Gaze} {Estimation}. In: 2022 {IEEE}/{CVF} {Conference} on {Computer} {Vision} and {Pattern} {Recognition} ({CVPR}). pp. 19354--19363. IEEE, New Orleans, LA, USA (Jun 2022)

\end{thebibliography}
%




\end{document}


\title{Contribution Title}
%
%


\section*{Supplemental materials for \\
Enhancing quantitative image synthesis through pretraining and resolution scaling for bone mineral density estimation from a plain X-ray image}

\begin{figure}
\centering
\includegraphics[width=\textwidth]{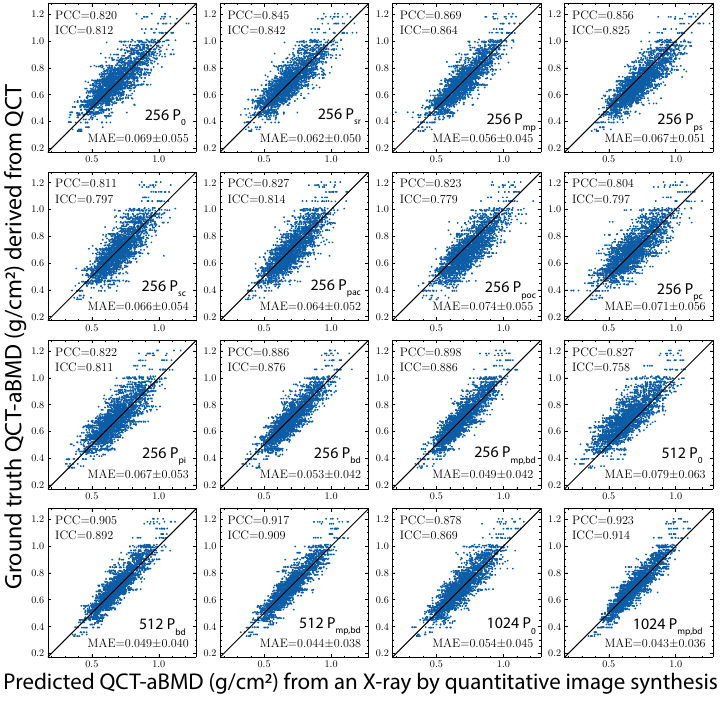}
\caption{The scatter plot of predicted QCT-aBMD against ground-truth QCT-aBMD for various pretrained methods (denoted $\mathbf{P}_{i}$) with different resolutions (denoted by 256, 512, and 1024) experimented on a 2656-X-ray dataset using four-fold cross-validation.
The predicted QCT-aBMD was calculated from a synthesized quantitative image from a plain X-ray image.
} \label{fig:bmd_scatter}
\end{figure}

\begin{table}[]
\centering
\caption{Training parameters for pretraining and target training.
NFNet-F0 \cite{brock_high-performance_2021} was used as the encoder.
The decoder followed \cite{gu_bone_2023}.
AdamW \cite{loshchilov_decoupled_2019} with a start learning rate of $1\times10^{-4}$ and weight decay of $1\times10^{-4}$ and the SGDR ($T_{0}=10$, $T_{mul}$=2) \cite{loshchilov_sgdr_2017} learning rate scheduler were used for all the training. 
The $\lambda_{L2}$ and $\lambda_{GC}$ were set to 10 and 1, respectively, for the target training.
}
\begin{tabular}{ccccccc}
\hline
                            & \multicolumn{1}{c|}{}                       & \multicolumn{3}{c|}{Pretraining {[}base{]}}                                                                       & \multicolumn{2}{c}{Target training}                         \\
\multirow{-2}{*}{Res.}      & \multicolumn{1}{c|}{\multirow{-2}{*}{Exp.}} & Batch size                       & Epoch                      & \multicolumn{1}{c|}{Other}                        & Batch size               & Epoch                     \\ \hline
256                         & $\mathbf{P_{0}}$                                          & /                                & /                          & /                                                 & 8                        & 70                        \\
{\color[HTML]{C0C0C0} 256}  & $\mathbf{P_{sr}}$                                           & 8                                & 1270                       & /                                                 & {\color[HTML]{C0C0C0} 8} & {\color[HTML]{C0C0C0} 70} \\
{\color[HTML]{C0C0C0} 256}  & $\mathbf{P_{mp}}$                                           & {\color[HTML]{C0C0C0} 8}         & 630                        & $8\times16$ patches                                      & {\color[HTML]{C0C0C0} 8} & {\color[HTML]{C0C0C0} 70} \\
{\color[HTML]{C0C0C0} 256}  & $\mathbf{P_{ps}}$                                           & {\color[HTML]{C0C0C0} 8}         & 1270                       & $4\times8$ patches                                      & {\color[HTML]{C0C0C0} 8} & {\color[HTML]{C0C0C0} 70} \\
{\color[HTML]{C0C0C0} 256}  & $\mathbf{P_{sc}}$                                           & {\color[HTML]{C0C0C0} 8}         & 630                        & Feat dim=2048, MoCo K=4096                        & {\color[HTML]{C0C0C0} 8} & {\color[HTML]{C0C0C0} 70} \\
{\color[HTML]{C0C0C0} 256}  & $\mathbf{P_{pac}}$                                           & {\color[HTML]{C0C0C0} 8}         & {\color[HTML]{C0C0C0} 630} & {\color[HTML]{C0C0C0} Feat dim=2048, MoCo K=4096} & {\color[HTML]{C0C0C0} 8} & {\color[HTML]{C0C0C0} 70} \\
{\color[HTML]{C0C0C0} 256}  & $\mathbf{P_{poc}}$                                           & {\color[HTML]{C0C0C0} 8}         & {\color[HTML]{C0C0C0} 630} & {\color[HTML]{C0C0C0} Feat dim=2048, MoCo K=4096} & {\color[HTML]{C0C0C0} 8} & {\color[HTML]{C0C0C0} 70} \\
{\color[HTML]{C0C0C0} 256}  & $\mathbf{P_{pc}}$                                           & {\color[HTML]{C0C0C0} 8}         & 1270                       & /                                                 & {\color[HTML]{C0C0C0} 8} & {\color[HTML]{C0C0C0} 70} \\
{\color[HTML]{C0C0C0} 256}  & $\mathbf{P_{pi}}$                                           & {\color[HTML]{C0C0C0} 8}         & 630                        & /                                                 & {\color[HTML]{C0C0C0} 8} & {\color[HTML]{C0C0C0} 70} \\
{\color[HTML]{C0C0C0} 256}  & $\mathbf{P_{bd}}$                                           & {\color[HTML]{C0C0C0} 8}         & 150                        & /                                                 & {\color[HTML]{C0C0C0} 8} & {\color[HTML]{C0C0C0} 70} \\
{\color[HTML]{C0C0C0} 256}  & $\mathbf{P_{mp,bd}}$                                          & {\color[HTML]{C0C0C0} 8} {[}{\color[HTML]{C0C0C0} 8}{]} & {\color[HTML]{C0C0C0} 150} {[}630{]}              & / {[}$8\times16$ patches{]}                              & {\color[HTML]{C0C0C0} 8} & {\color[HTML]{C0C0C0} 70} \\ \hline
512                         & $\mathbf{P_{0}}$                                           & /                                & /                          & /                                                 & 4                        & {\color[HTML]{C0C0C0} 70} \\
{\color[HTML]{C0C0C0} 512}  & $\mathbf{P_{bd}}$                                           & 4                                & 150                        & /                                                 & {\color[HTML]{C0C0C0} 4} & {\color[HTML]{C0C0C0} 70} \\
{\color[HTML]{C0C0C0} 512}  & $\mathbf{P_{mp,bd}}$                                          & {\color[HTML]{C0C0C0} 4} {[}{\color[HTML]{C0C0C0} 4}{]} & {\color[HTML]{C0C0C0} 150} {[}630{]}              & / {[}$8\times16$ patches{]}                              & {\color[HTML]{C0C0C0} 4} & {\color[HTML]{C0C0C0} 70} \\ \hline
1024                        & $\mathbf{P_{0}}$                                           & /                                & /                          & /                                                 & 2                        & {\color[HTML]{C0C0C0} 70} \\
{\color[HTML]{C0C0C0} 1024} & $\mathbf{P_{mp,bd}}$                                          & 2 {[}2{]}                        & 30 {[}630{]}               & / {[}$8\times16$ patches{]}                              & {\color[HTML]{C0C0C0} 2} & {\color[HTML]{C0C0C0} 70} \\ \hline
\end{tabular}
\end{table}

\bibliographystyle{splncs04}
\bibliography{Paper-030}

%



